\documentclass[a4,%
amsmath,amssymb,
12pt
]{revtex4-1}

\usepackage{bm}
\usepackage[T2A]{fontenc}
\usepackage{graphicx}
\usepackage{caption}

\begin{document}
	
	\title{Irreversibility in two-body system. Toy model} 
	
	\author{A. Yu. Zakharov}\email[E-mail: ]{Anatoly.Zakharov@novsu.ru; A.Yu.Zakharov@gmail.com} 
	
	\affiliation{Yaroslav-the-Wise Novgorod State University \\ 41, B. S.-Peterburgskaya Str., Veliky Novgorod 173003, Russia} 
	
	\begin{abstract}
		It is shown that the irreversible behavior of classical systems of interacting particles is a common property for both few-body and many-body systems. It is due to the delay in the interactions between the particles.
		
		\pacs: 45.50.Jf Few- and many-body systems; 
		
		82.40.Bj Oscillations, chaos, and bifurcations. 
		
		05.70.-a Thermodynamics

		
	\end{abstract}
	\maketitle

\section{Introduction}

It is well known that all the field theories have a property of the interactions retardation. Such are the electromagnetic interaction of charges in vacuum, the interaction of vibrational systems through a medium, the interactions between the particles in a plasma or colloidal systems, etc. The interactions delay leads to a violation of Newton's third law on the equality of action and reaction~\cite {Ivlev}, and also it is one of the mechanisms for the irreversible behavior of many-body systems~\cite{Zakh2,Zakh3}. 

Note that in papers~\cite{Zakh2,Zakh3} neither probabilistic considerations nor the conditions $N \gg 1$ ($N$ is number of degrees of freedom in the system) were used. In this connection, it is of interest to study in detail the mechanisms of the manifestation of retardation in the irreversibility of systems with a small number of degrees of freedom.

We define a two-body oscillator as a system of two particles with identical masses $ m $, the interaction between which \textbf {at rest} is described by the potential $ W \left (\mathbf {R} _1 - \mathbf {R} _2 \right) $ with the following properties.
\begin{enumerate}
	\item The function $W\left(\mathbf{R}_1 - \mathbf{R}_2 \right)$ has a minimum at  $\left|\mathbf{R}_1 - \mathbf{R}_2 \right| = L$.
	\item Near the minimum point, this function can be approximated by a quadratic function
	\begin{equation}\label{kx2}
	W\left(\mathbf{R}_1 - \mathbf{R}_2 \right) = W\left(L \right) + \frac{k\left( \mathbf{R}_1 - \mathbf{R}_2 \right) ^2}{2}.
	\end{equation}
\end{enumerate}
Let us consider the case of one-dimensional \textbf{small oscillations} of these particles along a straight line connecting these bodies.

In the absence of retardation, the solution of this problem is trivial and describes the oscillations of the particles with circular frequency $\displaystyle \sqrt{\frac{2k}{m}} $.
The frequency of these oscillations does not depend on either $L$ or $W\left(L \right) $. The retardation in the interactions between these two particles leads to a qualitatively different result, including the irreversible behavior of the system as a whole.

\section{Model with retardation}

We denote the deviations from the equilibrium positions of the particles in the rest state by $x_1\left( t\right) $ and $x_2\left( t\right) $. 
Equations of motion for the system in view of the interaction retardation~$\tau$  have the form: 
\begin{equation}\label{Eq-2}
\left\lbrace 
\begin{array}{l}
{\displaystyle \ddot{x}_1(t)\, + \omega_0^2 \left[x_1(t) - x_2(t-\tau) \right] = 0; } \\
{\displaystyle \ddot{x}_2(t)\, + \omega_0^2 \left[x_2(t) - x_1(t-\tau) \right] = 0,}
\end{array}
\right.
\end{equation}
where $\omega_0 = \sqrt{\frac{k}{m}}$. 

In general case, $\tau$ is not a constant, but a function of the particles positions, that is, of the solution of the system of equations~(\ref{Eq-2}). The solution of such a problem in a general formulation goes far beyond the possibilities of modern mathematics.
The condition of smallness of the oscillations is 
\begin{equation}\label{small}
\left| x_k\left(t \right) \right| \ll L, \quad (k=1, 2). 
\end{equation} 
It substantially simplifies the situation since in this case the principal term of the interaction retardation is reduced to the constant
\begin{equation}\label{tau}
\tau = \frac{L}{v},
\end{equation}
where $v$ is the propagation velocity of the interaction. Thus, in the case of small oscillations, the system of equations~(\ref{Eq-2}) is linear.

The Euler substitution 
\begin{equation}\label{Euler}
x_k\left(t \right)  = C_k\, e^{i\, \Omega t}
\end{equation}
leads to the following characteristic equation with respect to~$ \Omega $:
\begin{equation}\label{char2}
\Delta \left(\Omega \right) = \left| 
\begin{array}{cc}
\left( \omega_0 ^2 - \Omega^2\right) & \quad - \omega_0 ^2 e^{-i\Omega\tau}\\
- \omega_0 ^2  e^{-i\Omega\tau} & \quad \left( \omega_0 ^2 - \Omega^2\right)
\end{array}
\right| =   \Omega^4 -2 \omega_0^2\,\Omega^2 +\omega_0^4\left[1-e^{-2i\Omega\tau} \right] = 0.
\end{equation}

The roots of this transcendental equation depend on~$ \tau $ and in general case are complex:
\begin{equation}\label{Im-alpha}
\Omega = \omega +i\gamma.
\end{equation}
The equation~(\ref {char2}) with respect to the \textbf {complex} value~$ \Omega $ is equivalent to a system of two equations for \textbf{real} quantities~$ \omega $ and $ \gamma $:
\begin{eqnarray}
{\displaystyle  \omega^4 - 6\omega^2 \gamma^2+\gamma^4- 2 \omega_0^2\left( \omega^2 - \gamma^2\right) +\omega_0^4\left[1-e^{2\gamma\tau} \cos{2\omega\tau }\right]  =0; }\label{ab} \\
{\displaystyle 4\omega \gamma\left(\omega^2-\gamma^2 \right) -4\omega_0^2 \omega \gamma + \omega_0^4\, e^{2\gamma\tau } \sin{2\omega\tau }=0.}\label{ba}
\end{eqnarray}
Note that $ \omega $ determines the oscillation frequency, and $ \gamma $ characterizes the rate of change in the amplitude of the oscillations. Therefore, the condition for stationarity of the oscillations is that $ \gamma=0 $, whence we get 
\begin{equation}\label{b=0}
2\omega\tau = \frac{2\omega L}{v} = \pi n,
\end{equation}
where $n$~is an arbitrary natural number. 

Substituting $ \tau $ from~(\ref {b=0}) and $ \gamma = 0 $ into equation~(\ref{ab}), we obtain the equation for~$ \omega $
\begin{equation}\label{a-n}
\omega^4 - 2 \omega_0^2\, \omega^2 +2\omega_0^4\,\sin^2\left(\frac{\pi n}{2} \right) =0.
\end{equation}
The roots of this equation are real if and only if $ n $ is an even number. Under this condition, the solutions are as follows:
\begin{equation}\label{a1-a4}
\omega_1=\omega_2=0; \qquad \omega_{3,4}=\pm\sqrt{2}\,\omega_0.
\end{equation}
Thus, stationary oscillations of a two-particle oscillator with retarded interactions occur only for a discrete set of equilibrium distances~$ L $ between the particles, determined by the condition~(\ref {b=0}).

However, there exists a set of values of the parameter~$ L $ for which both stationary and non-stationary oscillations are possible, i.e. solution of the system of equations~(\ref {ab})--(\ref{ba}) with $\gamma = 0 $ and $\gamma \not = 0 $, respectively. 
From the immense set of solutions of this system of equations depending on the parameter~$ L $, we consider a subset for which the condition~(\ref {b=0}) is satisfied:
\begin{equation}\label{sin=0}
L=\frac{\pi n v}{2\omega}.
\end{equation} 
In this case, the system of equations~~(\ref {ab})--(\ref{ba}) is greatly simplified:
\begin{equation}\label{a-b-2}
\left\lbrace 
\begin{array}{l}
{\displaystyle \omega^4 - 6\omega^2\gamma^2+\gamma^4- 2 \omega_0^2\left( \omega^2 - \gamma^2\right) +\omega_0^4\left[1-e^{2\gamma\tau} \left( -1\right)^n  \right]  =0; }\\
{\displaystyle 4\omega \gamma\left(\omega^2-\gamma^2 - \omega_0^2\right) = 0.}
\end{array}
\right. 
\end{equation}

For $ \gamma \not = 0 $ we have
\begin{equation}\label{b-nonzero}
\gamma=\eta \sqrt{\omega^2 - \omega_0^2},
\end{equation}
where $\eta=\pm 1$.

Substituting this expression for~$ \gamma $ into the first of the equations~(\ref {a-b-2}), we find
\begin{equation}\label{a-2}
\left( -1\right)^n\, e^{2\gamma\tau} =   4 \left( \frac{\omega}{\omega_0}\right) ^2 \left(1 - \left( \frac{\omega}{\omega_0}\right) ^2 \right). 
\end{equation}

As follows from the domain of definition of the expression~(\ref {b-nonzero}) for $ \gamma $, the right-hand side of the equation~(\ref {a-2}) is non-positive, so $ n $ in this equation is an odd number ($ n = 2s + 1 $), and the equation itself becomes
\begin{equation}\label{ln}
(2s+1)\eta  = \frac{1}{\pi}\ \frac{x\,\ln \left[4\left( x^4 - x^2\right)  \right]}{\sqrt{x^2-1}}, 
\end{equation}
where $\displaystyle x^2= \left( \frac{\omega}{\omega_0}\right)^2 > 1$. 

The right-hand side of this equation is a monotonically increasing function on the interval $ (1, + \infty) $. The range of this function fills the entire interval $ (- \infty, + \infty) $. Therefore, for each value of $ \eta \left (2s + 1 \right) $, the equation (\ref {ln}) has an unique solution~$x_s(\eta)$.

\begin{figure}[b]
	\includegraphics[width=0.9\linewidth]{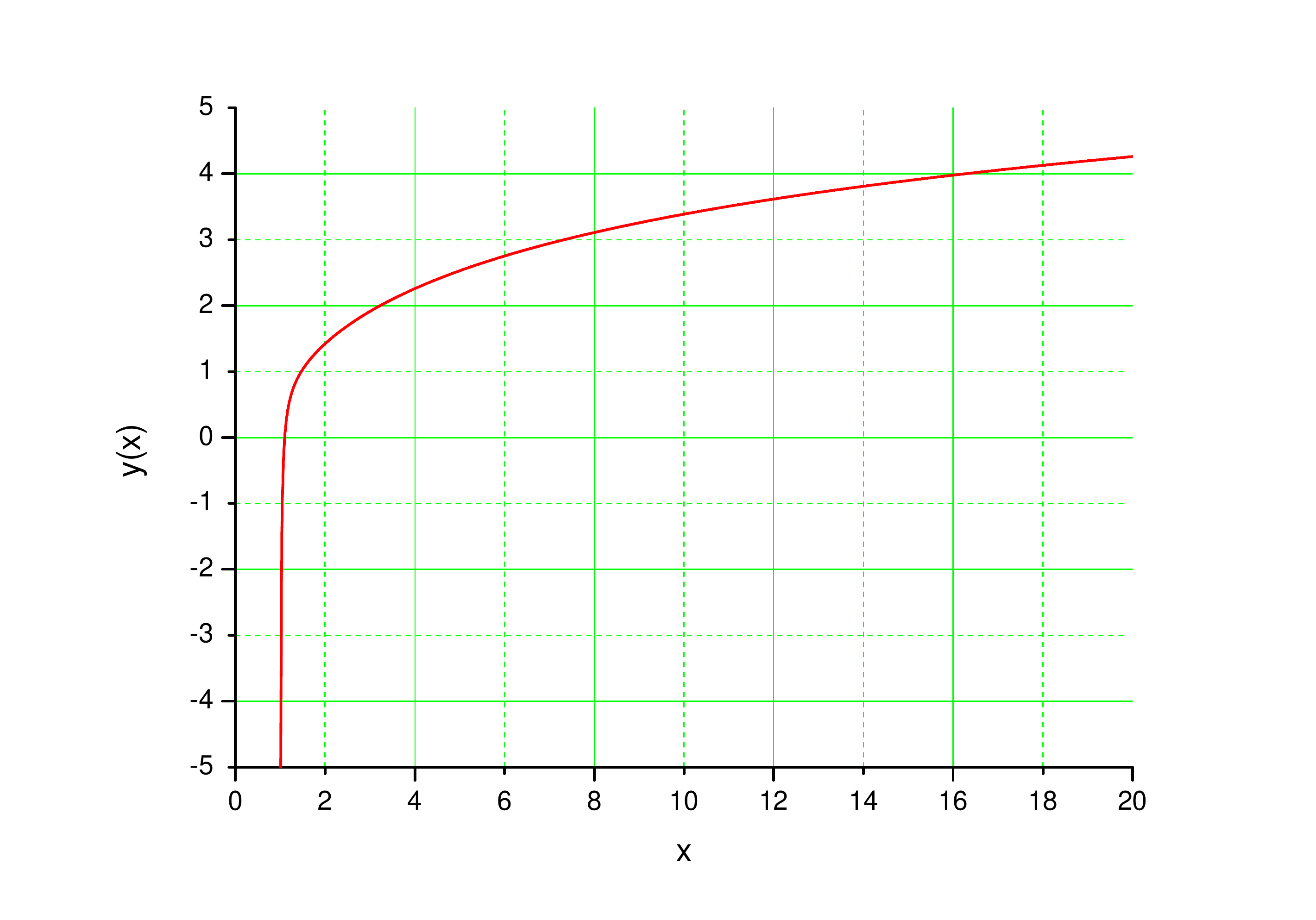}
	\caption{The graph of the function $y(x)=\frac{1}{\pi}\ \frac{x\,\ln \left[4\left( x^4 - x^2\right)  \right]}{\sqrt{x^2-1}}$.}
	\label{fig:oscillator}
\end{figure}

The graph of the function contained in the right-hand side of this equation is shown in~Fig.\ref{fig:oscillator}.

\section{Conclusion}

The delay in the interaction between constituents of two-body oscillator leads to the following effects:
\begin{enumerate}
	\item Stationary oscillations exist only for a discrete set of equilibrium distances between particles, determined by the condition~(\ref{b=0}). The frequencies of all stationary oscillations~$\omega$ are the same:
	\begin{equation}\label{freq-stat}
	\omega = \sqrt{2}\, \omega_0.
	\end{equation}	
	Condition~(\ref{b=0}), with account the relation~(\ref{freq-stat}), can be represented in an equivalent form
	\begin{equation}\label{b=0-2}
	L = \frac{n}{4} \lambda,
	\end{equation} 
	where $ \lambda = 2\pi v\sqrt{\frac{m}{2 k}}$ is the wavelength corresponding to the oscillation frequency~(\ref{freq-stat}).   
	It should be noted that the Lebesgue measure of the set of values of $L$ for which stationary oscillations of a two-particle oscillator exist is zero. 
	\item Under the same condition~(\ref{b=0}), there are infinite number of non-stationary oscillations with different frequencies corresponding to the complex roots of the characteristic equation~(\ref{char2}):
	\begin{equation}\label{char3}
	\omega_s\left(\eta \right)  = x_s(\eta )\, \omega_0, \quad \gamma_s\left( \eta\right)  =\eta\omega_0 \sqrt{x_s^2(\eta)-1}.
	\end{equation} 
	These non-stationary oscillations represent the irreversible behavior of a two-particle oscillator.
\end{enumerate}

The described simple example leads to the following conclusions.
\begin{enumerate}
	\item The irreversible behavior of systems of interacting particles is a common property for both few-body and many-body classical system, having its origin  in the delay of interactions. 
	\item The unavoidable delay of interactions is sufficient for the irreversible behavior of systems. The systems are irreversible in itself -- there is no need to use any probabilistic hypotheses or other assumptions to explain the phenomenon of irreversibility in systems of interacting particles. 
\end{enumerate}

\begin{acknowledgments}
I am grateful to Profs. Ya. I. Granovsky and V. V. Uchaikin for useful discussions.

The work is supported by the Ministry of Education and Science of the Russian Federation within the framework of the project part of the state order (project No.~3.3572.2017).
\end{acknowledgments}


\end{document}